\documentclass[namedreferences]{SolarPhysics}

\usepackage[optionalrh]{spr-sola-addons} 
\usepackage{graphicx}        
\usepackage{color}           
\usepackage{url}             

\newcommand{\etal}{{\it et al.}}

\begin{document}

\begin{article}

\begin{opening}

\title{Heliospheric Transport of Neutron-Decay Protons}

\author{E.~E.~\surname{Chollet}$^{1}$\sep
R.~A.~\surname{Mewaldt}$^{1}$}

\runningauthor{Chollet \& Mewaldt}

\runningtitle{Solar Neutron-Decay Protons}

\institute{$^{1}$ California Institute of Technology, Mail Code 290-17, Pasadena, CA 91125 USA}

%
%

\begin{abstract} 
We report on new simulations of the transport of energetic protons originating from the decay 
of energetic neutrons produced in solar flares.  Because the neutrons are fast-moving but 
insensitive to the solar wind magnetic field, the decay protons are produced over a wide 
region of space, and they should be detectable by current instruments over a broad range
of longitudes for many hours after a sufficiently large gamma-ray 
flare.  Spacecraft closer to the Sun are expected to see 
orders-of-magnitude higher intensities than those at the Earth-Sun distance.  
The current solar cycle should present an excellent opportunity to observe neutron-decay 
protons with multiple spacecraft over different heliographic longitudes and distances 
from the Sun.
\end{abstract}
\keywords{Flares, Energetic Particles; Energetic Particles, Neutrons;
Energetic Particles, Protons; Energetic Particles, Propagation}
\end{opening}

\section{Introduction}

Solar flares remain some of the best laboratories in the solar system for studying particle 
acceleration, production and interaction.  The particles accelerated in flares either 
interact with the solar chromosphere, producing observable X-ray and $\gamma$-ray emission, 
or escape into the interplanetary medium.  In 1951, Biermann, Haxel and Schluter first 
suggested that solar flares could produce energetic neutrons, and evidence of these 
neutrons was subsequently observed in the 2.2-MeV neutron-capture $\gamma$-ray line, 
first by \inlinecite{chu73} and many times since.  The first direct observations of 
solar-flare neutrons were made by \inlinecite{chu82}.  Detailed models of neutron 
production, transport, interaction and decay in flares have been 
developed ({\it e.g.} \opencite{lin65}, \opencite{hua87}, \opencite{mur07}), allowing the 
number of interacting protons and escaping neutrons from a particular flare to be 
deduced from the 2.2 MeV line fluence.  The neutrons which escape the Sun undergo 
beta decay with a mean lifetime at rest of 886 seconds (about 14 minutes):

\begin{equation}
n^{0}~\longrightarrow~p^{+}+e^{-}+\overline{\nu}_{e}
\end{equation}
creating neutron-decay protons and electrons in the heliosphere.

The escaping neutrons and their decay products have been observed by both ground-based 
detectors and space-based instruments near the Earth and around the heliosphere 
(\opencite{chu82}, \opencite{deb83}).  Since the travel time between the Sun and Earth 
is $\geq$20 minutes for particles with energies less than 100 MeV, most of the neutrons 
decay at some time during transit to 1 AU.  The neutron-decay protons \cite{eve83a, eve90} 
and also electrons \cite{dro96} typically arrive before the 
original flare-accelerated protons (protons of the same energy accelerated in the 
flare site rather than originating from neutron decay), and before particles 
accelerated by the shock driven by 
the associated coronal mass ejection (CME), because they 
make part of their journey as neutrons that travel in a straight line and also 
do not lose energy.

Following the early work by Roelof (1966), several studies have
investigated the effects  that transport in the interplanetary 
magnetic field (IMF) can have on observations of neutron-decay protons, including 
their time-intensity profiles, energy spectra, and pitch-angle distributions (PADs). 
The first measurements of neutron-decay protons found the PADs were nearly isotropic, 
and, by fitting various aspects of the data, obtained scattering mean free-paths of 
$\lambda$ = 0.3--0.5 AU \cite{eve83a}, $\lambda$ = 0.2 $\pm$ 0.005 AU \cite{eve83b}, 
and $\lambda\sim$ 0.3 AU \cite{eve85}.  In a more complete treatment \inlinecite{ruf91} 
modeled the transport of neutron-decay protons with 25 to 150 MeV and fit the data from events summarized by 
\inlinecite{eve85}.  He considered pitch-angle scattering and adiabatic focusing of the 
protons and, by fitting ``distance-traveled'' distributions deduced from data obtained by the {\it Third International Sun-Earth Explorer} (ISEE-3), 
obtained the best-fit scattering mean free-paths for two events of $\lambda$ = 0.26 $\pm$ 
0.05 AU and $\lambda$ = 0.37 $\pm$ 0.06 AU.  Recently, \inlinecite{agu11} modeled 
the transport of neutrons and their decay products to compare with observations of 
Solar Cycle 23 events.

Previous {\it in situ} studies of neutrons and their decay products were limited to a
single vantage point, but new multi-point instrumentation is providing a 
different perspective on 
the geometry of energetic particles in the solar system.  The canonical 
view \cite{rea91,nit06} holds that CME shocks 
distribute particles widely in longitude, and flare-associated protons and 
electrons cover a narrow ($\sim$40$^\circ$) range in longitude.  However, 
recent studies have found that small flares can produce particles over 
$\sim$80$^\circ$ latitude \cite{wib06, wie09}, while new imaging technology 
is routinely showing previously-undetectable CMEs associated with small flares.  
Understanding the longitude-distribution of neutron-decay protons may help unravel 
the diverse sources of energetic particles in the heliosphere.  The twin spacecraft 
of the {\it Solar TErrestrial RElations Observatory} (STEREO, \opencite{kai08}) were launched 
in 2006 into orbits that lead (STEREO-Ahead) and trail (STEREO-Behind) the Earth.  
The {\it Advanced Composition Explorer} (ACE, \opencite{sto98}), {\it Solar and Heliospheric 
Observatory} (SOHO, \opencite{dom83}), and {\it Wind} spacecraft \cite{acu95} are in
orbit around Earth's L1 Lagrangian point, making multi-point studies covering 
more than 180$^\circ$ in longitude possible in the current solar cycle.

\section{Description of the Model}

To better understand what these spacecraft should observe in 
neutron-decay protons, we have created a model of neutron and proton transport 
in the heliosphere which determines the trajectories of a large number of 
individual particles by integrating their equations of motion.  Neutrons are 
initially injected as a delta function in space and time at the solar surface, 
for simplicity set at 0$^\circ$ latitude and longitude. The energy spectrum and 
angular distribution of the injected neutrons follow two of the examples that were 
investigated by 
\inlinecite{mur07}, and illustrated in their Figure 15, where the 
flare-accelerated proton spectrum was assumed to be a power law with a spectral 
index of -4, and the accelerated composition was characterizedby He/H = 0.5 along 
with corresponding enrichments in the other species with atomic number Z$>$2.  

\inlinecite{mur07} varied two additional parameters: the mean free-path for 
pitch-angle scattering (PAS) of accelerated protons in the solar atmosphere 
(which we label as $\lambda_{s}$ to distinguish it from the scattering mean 
free-path in interplanetary space, which we label as $\lambda$) and the magnetic 
convergence parameter ($\delta$). In \inlinecite{mur07} $\lambda_{s}$ is a 
dimensionless ratio of a) the distance for an arbitrary initial angular distribution 
to isotropize to b) the half-length of the loop on which the protons are trapped.  
They considered cases of no PAS ($\lambda_{s}\to\infty$), intermediate PAS 
($\lambda_{s}$ = 300), and high scattering ($\lambda_{s}$ = 20). The magnetic 
convergence parameter characterizes the magnetic field strength below the 
transition region, which is assumed to be proportional to the pressure to the 
power $\delta$. For $\delta$ = 0 there is no mirroring, and the angular
distribution of the interacting protons is downward-isotropic. For $\delta$ = 0.20 the 
protons will mirror as the field increases with depth, leading to an angular 
distribution of escaping neutrons that is somewhat peaked at 90$^{\circ}$ to 
the zenith, especially at energies $>$10 MeV. 

In this paper we consider two examples:  Case 1 (where $\lambda_{s}$ = 300), 
and Case 2 (with $\lambda_{s}\to\infty$), both with $\delta$ = 0.2. The corresponding 
spectra of escaping neutrons are shown in Figure 15 of \inlinecite{mur07} in 
three angular ranges.  We have approximated the escaping neutron spectra
with piece-wise power laws covering three energy intervals and three angular 
ranges from $\cos\theta$ = 0$^{\circ}$ to $\cos\theta$ = 90$^{\circ}$ as 
specified in Tables 1 and 2. Here $\theta$ is the angle relative to the surface 
normal, with $\theta$ = 0$^{\circ}$ being the direction away from the Sun. We have considered 
neutrons with initial energies between 0.1 and 100 MeV, with flat spectra 
between 0.1 and 1 MeV based on a projection of the curves in Figure 15 of \inlinecite{mur07}.  
Because our code does not include relativistic corrections to the energy loss during 
transport, we do not consider particles with energies greater than 100 MeV.

Calculations based on observed 2.2-MeV line 
fluences have shown that in large flares between $\sim$3x10$^{28}$ and 
$\sim$10$^{33}$ protons with energies $>$30 MeV are accelerated \cite{shi09}, 
and the calculations of neutron production described above show that for
the Case 1 spectra about 0.0054 escaping neutrons are produced for each 
accelerated $>$30 MeV proton \cite{mur07}. For the Case 2 spectra where
$\lambda_{s}\to\infty$ about 0.0073 escaping neutrons are produced
for each accelerated proton $>$30 MeV.  The yields in Tables 1 and 2 are normalized to 
1 flare-accelerated proton $>$30 MeV.

\begin{table}
\caption{Representation of the Escaping Neutron Spectra: Case 1$^{1}$}
\begin{tabular}{cccccccc} 
\hline
Energy & \multicolumn{2}{c}{$\cos\theta$ = 0.0--0.3} & \multicolumn{2}{c}{$\cos\theta$ = 0.3--0.7} & \multicolumn{2}{c}{$\cos\theta$ = 0.7--1.0} & $\cos\theta\geq 0$ \\
(MeV) & Intercept$^{2}$ & Slope$^{2}$ & Intercept & Slope & Intercept & Slope & N/P30$^{3}$ \\
\hline
0.1--1 & 4.51E-5 & 0.00 & 6.85E-5 & 0.00 & 7.05E-5 & 0.00 & 3.51E-4 \\
1--10 & 4.67E-5 & -0.223 & 7.83E-5 & -0.429 & 9.77E-5 & -0.830 & 2.02E-3 \\
10--100 & 5.10E-4 & -1.16 & 6.13E-4 & -1.30 & 8.95E-4 & -1.68 & 2.99E-3 \\ \hline 
0.1--100 & & & & & & & \\
N/P30$^{3}$ & \multicolumn{2}{c}{1.94E-3} & \multicolumn{2}{c}{2.38E-3} & \multicolumn{2}{c}{1.04E-3} & 5.36E-3 \\
\hline\\ 
\end{tabular}
\noindent$^{1}$Case 1 is based on Figure 15b of \cite{mur07} with $\lambda_{s}$ = 300 and $\delta$ = 0.20 (see text)\\
\noindent$^{2}$The Intercept and Slope entries define power-law representations of neutron energy spectra for the indicated energy and angular ranges.\\
\noindent$^{3}$N/P30 is the number of neutrons leaving the Sun for each accelerated proton $>$30 MeV. 
N/P30 values in the last column and row are integrated over angle and energy respectively.\\
\end{table}

\begin{table}
\caption{Representation of the Escaping Neutron Spectra: Case 2$^{1}$}
\begin{tabular}{cccccccc} 
\hline
Energy & \multicolumn{2}{c}{$\cos\theta$ = 0.0--0.3} & \multicolumn{2}{c}{$\cos\theta$ = 0.3--0.7} & \multicolumn{2}{c}{$\cos\theta$ = 0.7--1.0} & $\cos\theta\geq 0$ \\
(MeV) & Intercept$^{2}$ & Slope$^{2}$ & Intercept & Slope & Intercept & Slope & N/P30$^{3}$ \\
\hline
0.1--1 & 5.42E-5 & 0.00 & 6.78E-5 & 0.00 & 8.33E-5 & 0.00 & 3.86E-4 \\
1--10 & 5.85E-5 & -0.112 & 8.46E-5 & -0.422 & 9.00E-5 & -0.843 & 2.33E-3 \\
10--100 & 7.34E-4 & -1.10 & 6.58E-4 & -1.22 & 7.34E-4 & -1.52 & 4.58E-3 \\ \hline 
0.1--100 & & & & & & & \\
N/P30$^{3}$ & \multicolumn{2}{c}{3.17E-3} & \multicolumn{2}{c}{2.95E-3} & \multicolumn{2}{c}{1.17E-3} & 7.29E-3 \\
\hline\\ 
\end{tabular}
\noindent$^{1}$Case 2 is based on Figure 15a of \cite{mur07} with $\lambda_{s}\to\infty$ and $\delta$ = 0.20 (see text)\\  
\noindent$^{2}$The Intercept and Slope entries define power-law representations of neutron energy spectra for the indicated energy and angular ranges.\\
\noindent$^{3}$N/P30 is the number of neutrons leaving the Sun for each accelerated proton $>$30 MeV. 
N/P30 values in the last column and row are integrated over angle and energy respectively.\\
\end{table}

At each timestep in the simulation, the escaping neutrons are tested for decay by 
comparing a random number between 0 and 1 with a decay probability given by 
$P_d = \Delta t/\tau_{d}$ where $\tau_{d}$ is the mean lifetime (corrected 
for time dilation). Once a neutron decays, the trajectory of the resulting proton is 
followed as it begins to feel the effects of the interplanetary magnetic 
field, assumed to be a Parker spiral for a solar wind speed of 400 km s$^{-1}$.  
Rather than computing the gyromotion directly, we treat the particles adiabatically, 
that is, we assume the perpendicular kinetic energy over the magnetic field 
magnitude ($W_\perp/B$) is constant in the frame corotating with the Sun.  
This scheme produces results in excellent agreement with schemes where the 
gyromotion is integrated directly, but 
results in a significant savings of computational resources. The proton 
trajectory is followed until it is lost to the Sun, passes the outer boundary 
at 10 astronomical units (AU, 1 AU is the Earth-Sun distance), or until it has 
been in transit for 10 days.   

Our code includes the effects of pitch-angle scattering and adiabatic cooling 
of the protons in the expanding solar wind.  The particle scattering is included 
in a manner similar to the particle decay, where a scatter probability is compared 
to a random number for each timestep. The form of the scattering mean free-path 
is derived from the quasilinear theory \cite{jok66,gia99}:
\begin{equation}
\lambda_{QLT}(r,v) = 0.1\bigg({v\over v_o}\bigg)^{1\over
3}\bigg({B(1AU)\over B(r)}\bigg)^{1\over 3},
\end{equation}
where $r$ is radial distance, $v$ is particle velocity, $B$ is the strength 
of the IMF, and $v_o$ is the velocity of a 1-MeV proton. 

Our value for the scattering mean free-path $\lambda$ ranges from 0.17 to 
0.22 AU for 25--150 MeV protons at 1 AU and gradually decreases closer to the 
Sun, values that agree well with the previously observed PADs in neutron-decay 
proton events.  The recent study of neutron-decay protons by Agueda {\it et al.} (2011) 
assumed scatter-free propagation,  in which the decay-protons quickly focus 
into a beam aligned with the IMF.  In such a beam, protons cross a detector 
at 1 AU only once.  In our simulation, 10-MeV protons cross 1 AU an average 
of 6-7 times.  

In an earlier paper reporting on simulations of solar energetic particle (SEP) transport, 
\inlinecite{cho10} found that the proton energy deposition at 1 AU was constant 
to within a factor of two as the scattering mean free-path was varied from 
$\lambda$ = 0.01 to $\lambda$ = 1 AU.  It was also relatively insensitive 
to variations in the radial dependence of $\lambda$.  This is because when 
pitch-angle scattering is increased the protons cross 1 AU more times on 
average, but they also lose more energy (see Figure 7 in \inlinecite{cho10}).
If $\lambda$ is increased, we expect about the same proton fluence, but distributed 
over a shorter time period, with higher average intensity.
As a result, we believe the results of our simulations are relatively 
insensitive to the details of our description of pitch-angle scattering. 
The events observed to date suggest that the inclusion of some scattering 
scheme is necessary if the simulation is to match the
details of real events. 

Our simulation outputs the times particles cross a sphere at a given distance 
from the Sun along with their positions and energies, allowing intensities and 
fluences to be calculated at the simulated spacecraft. Calculations based on observed 
2.2-MeV line fluences have shown that in large flares between $\sim$3x10$^{28}$ 
and $\sim$10$^{33}$ protons with energies $>$30 MeV are accelerated \cite{shi09}. 

The calculations of neutron production described above show that for the Case 1 
spectra about 0.0054 escaping neutrons are produced for each accelerated $>$30 
MeV proton \cite{mur07}. For the Case 2 spectra where $\lambda_{s}\to\infty$ 
about 0.0073 escaping neutrons are produced for accelerated proton $>$30 MeV. 
Therefore, we have normalized the intensities and fluences to the Case 1 and 
Case 2 yields of 5.36 x 10$^{29}$ and 7.29 x 10$^{29}$ injected neutrons 
expected for 10$^{32}$ accelerated protons $>$30 MeV, somewhat greater than 
the median value observed in real flares. [At least 23\% of the 
{\it Ramaty High Energy Solar Spectroscopic Imager} (RHESSI) 
$\gamma$-ray flares in which the 2.2 MeV line was detected 
were this large \cite{shi09}; in some events RHESSI missed some 
of the emission because of shadowing by Earth.] Our total Case 2 yield 
of 1-100 MeV escaping neutrons is about a factor of $\sim$3.5 greater than 
for similar neutron spectra in \inlinecite{hua02}, which was the basis of 
the transport study by \inlinecite{agu11}. This is likely because Hua {\it et al.} 
assumed an accelerated composition less rich in He and heavier ions, as well 
as a harder proton spectrum (spectral index of -3.5).

\begin{figure}
\centerline{\includegraphics[width=\textwidth]{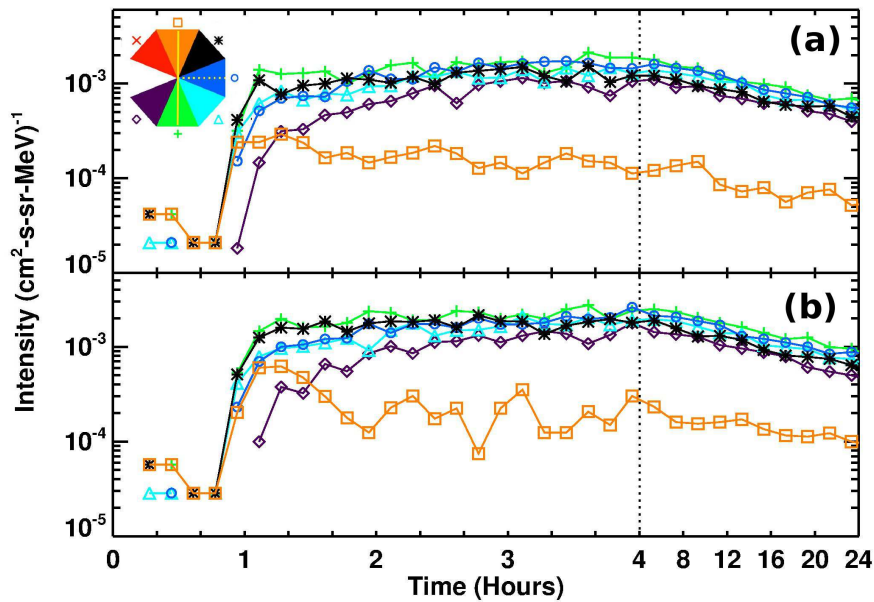}}
\caption{(a) Intensities of 1 to 10 MeV protons at 1 AU for spacecraft at 
different longitudes following a central-meridian
0$^{\circ}$-latitude $\gamma$-ray flare with 10$^{32}$ accelerated protons 
$>$30 MeV and Case-1 characteristics ($\lambda_{s}$ = 300 and $\delta$ = 0.20; see 
Murphy \etal\ (2007)). The time T = 0 marks the time the flare photons would be observed 
at Earth. To the left of the black dotted line, the intensities are binned in 
10-minute increments, and to the right of the 
dotted line the intensities are binned in 2-hour increments.  For all but the 
orange curve the mean statistical uncertainty is $\sim$18\% for the 10-minute points and 
$\sim$8\% for the 2-hour points. For the orange curve the mean statistical error
is $\sim$43\% for the 10-minute points and $\sim$19\% for the 2-hour points.  
The different colors and symbols represent the different virtual spacecraft, 
with a legend at the top left indicating the relative longitudes of the spacecraft.  
The dotted line in the legend is the radial direction from the injection point.  
The sector represented by the red x has intensities a factor of 10 lower than the 
adjacent sector, 
and so is off the scale in this plot. (b) Same as panel (a) but for Case 2 neutron spectra 
characterized by $\lambda_{s}\to\infty$ and $\delta$ = 0.20. The mean statistical 
uncertainties are similar to those in panel (a).  Both simulations used the same
sequence of random numbers.}
\end{figure}

\section{Results and Discussion}

Figure 1 shows Case 1 and Case 2 intensities of energetic protons vs. time for 
a simulated spacecraft at 1 AU located at different heliographic longitudes.  
Each of the seven simulated spacecraft integrate over 45$^\circ$ in longitude 
and 15$^\circ$ in latitude.  The 1-AU intensities reflect the greater yield of 
escaping neutrons in Case 2, in most directions the intensities are somewhat 
higher in Figure 1b than in Figure 1a. Increasing the amount of PAS decreases 
the yield of escaping neutrons because more accelerated protons are scattered 
into the loss cone to the chromosphere, so that they interact deeper in the 
atmosphere while typically headed downward. Since the secondary neutrons generally 
go forward at high energy, fewer neutrons escape. When PAS is ignored, more neutrons 
interact while mirroring and so there is increased neutron escape near the solar 
horizon ($\theta$ = 90$^\circ$).
 
The intensities in Figure 1 are above the background levels of some currently-available 
instruments over more than half of the longitudinal circle for most of a day. 
The {\it Low Energy Telescope} (LET) and {\it High Energy Telescope} (HET) sensors on STEREO, 
which observe protons between 2 and 100 MeV, have a galactic cosmic ray background 
level of a few times $10^{-5}$ (cm$^{2}$-s-sr-MeV)$^{-1}$ at solar maximum 
(see also \opencite{gom10}, \opencite{les08}, and \opencite{mew09}). With this 
instrument background level, the simulated event would be detectable at five 
of the virtual spacecraft in this simulation (225$^{\circ}$) for a few hours to a day. 
The estimated decay-proton intensities for the 28 October 2003 ``Halloween'' event 
would be $\sim$10 times greater. \inlinecite{agu11} conducted a survey of 1 to 11 
MeV {\it Wind-3DP} data for eight X-class flares, and found no evidence for detection of 
neutron-decay protons.  However, their background levels were $\sim$0.04-0.1 
cm$^{-2}$ sr$^{-1}$ MeV$^{-1}$ for 3-6 MeV protons and $\sim$0.02 cm$^{-2}$ sr$^{-1}$ 
MeV$^{-1}$ for 6-11 MeV protons, 
and they note that their simulated neutron-decay proton spectrum is well below the 
observed background proton spectrum for the only one of these events in 
which RHESSI observed 2.2-MeV $\gamma$-ray emission.

Though the intensities predicted by our simulation are above instrument detection 
thresholds, for magnetically well-connected events it is difficult to distinguish 
neutron-decay protons from shock and flare accelerated protons \cite{roe66}. Events 
originating in the eastern hemisphere are more likely to be detected because the 
directly-accelerated protons often arrive several hours after the neutron-decay products.  
However, many of the largest SEP events are preceded by earlier events that elevate 
interplanetary SEP levels, which may explain why so few detections of neutron-decay 
events have been reported.

In Figure 2 we show snapshots of particle positions in the 
ecliptic plane 20 minutes and one hour after the flare, with the angles and symbols 
of the virtual spacecraft superimposed. 
This simulation shows that solar flares provide a means of quickly distributing 
neutron-decay protons (and electrons) over more than 180$^\circ$ in longitude within the first hour following the flare.
The solar surface creates a solid horizon that neutrons cannot cross. If the 
neutron-decay protons travel without scattering along interplanetary magnetic field 
lines, this horizon will create a shadowed region in 
interplanetary space as discussed by \inlinecite{eve83a} and \inlinecite{agu11}.  
The scattering included in this simulation allows protons to move against the 
solar wind flow into this shadowed region, albeit with difficulty. As a result, 
a spacecraft like {\it Messenger} that happens to be inside the shadow might still be 
able to detect neutron-decay protons that diffuse into the shadow region. 

\begin{figure}
\centerline{\includegraphics[width=\textwidth]{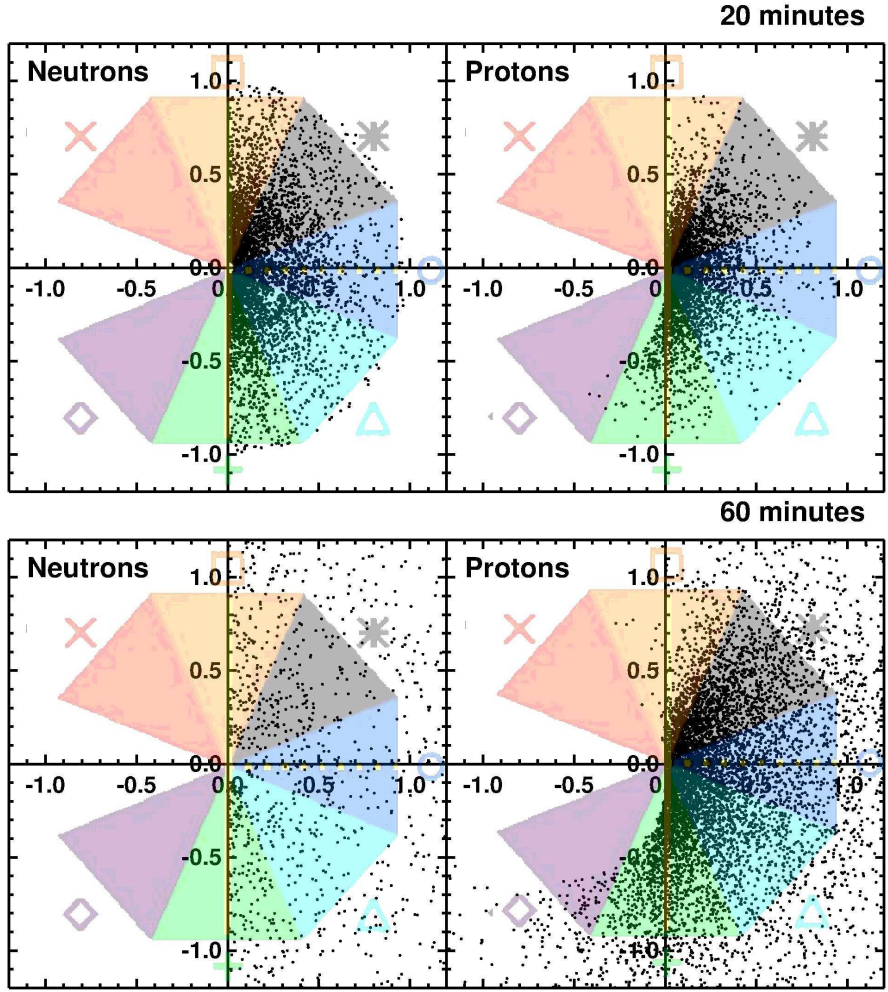}}
\caption{Snapshots of particle positions at 20 and 60 minutes 
after the neutrons are injected (in 20 minutes a 100 MeV neutron travels just
over 1 AU).  The left half shows the positions of the neutrons, while 
the right half shows the positions of neutron-decay protons.  A 
legend representing the sectors of the virtual spacecraft is 
superimposed.  The neutrons travel in straight lines, while the protons 
spiral while following the magnetic field.  This simulation used the Case-1
neutron spectra that includes PAS of the accelerated protons with 
$\lambda_{s}$ = 300 and $\delta$ = 0.20 (see text and Murphy \etal\ (2007)).}
\end{figure}

\inlinecite{mur07} found that more neutrons are 
produced close to the neutron horizon than in the radial direction (see Tables 
1 and 2, where the minimum neutron intensities are roughly normal to the flare 
site), and the interplanetary results reflect this. Under the influence of the 
interplanetary magnetic field neutron-decay protons move 
clockwise from the sector marked with a square (W67.5$^{\circ}$--W112.5$^{\circ}$) 
to the sector marked with an asterisk 
(W22.5$^{\circ}$--W67.5$^{\circ}$), and from one half of the sector 
marked with a plus (W45$^{\circ}$--W67.5$^{\circ}$) to the other 
(W22.5$^{\circ}$--W45$^{\circ}$), making the plus and asterisk 
sectors the highest intensities early in  the event (see also Figure 1).  Later in the 
event, the particles have moved even more clockwise into the diamond 
(E112.5$^{\circ}$--E157.5$^{\circ}$) and circle (E22.5$^{\circ}$--W22.5$^{\circ}$) 
sectors, while the square (W67.5$^{\circ}$--W112.5$^{\circ}$) sector quickly becomes 
depleted. 

Proton fluence spectra at 1 AU integrated over the first four hours following the 
neutron injection are presented for each sector in Figure 3. Most of the low-speed 
protons (with energies less than a few MeV) have not yet reached 1 AU at this time. 
The higher-speed protons reflect the injection spectra and transport effects. 
As mentioned above, \inlinecite{agu11} have simulated neutron-decay proton intensities 
at 1-AU using somewhat different neutron spectra and interplanetary transport conditions. 
Their intensity spectra are similar in shape to our fluence spectra (Figure 3), but our 
spectra show evidence for adiabatic energy loss below $\sim$10 MeV.  In addition, their 
predicted peak intensities are greater than ours, presumably because, without pitch-angle 
scattering, all of the protons of a given energy arrive within a narrow time interval.

\begin{figure}
   \centerline{\includegraphics[width=\textwidth]{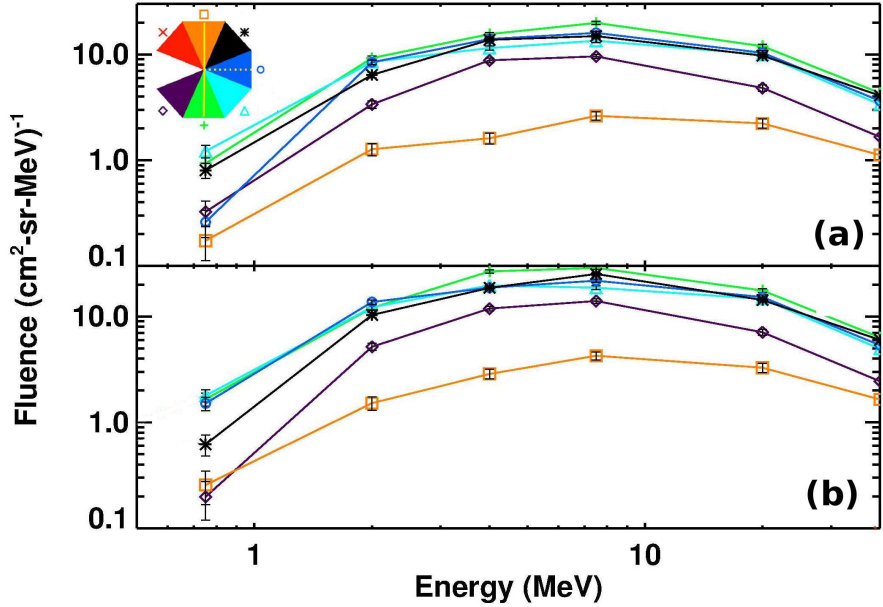}}
\caption{Proton fluence spectra integrated over the first four hours following neutron release 
(assumed to be the peak of the flare) are shown for each of the virtual spacecraft following a central-meridian 
$\gamma$-ray flare with 10$^{32}$ accelerated protons $>$30 MeV.  The relative intensities 
at the different spacecraft reflect the time-intensity profiles in Figure 1.  
Low-energy protons below 0.5 MeV have not yet reached 1 AU. Panel (a) is based on Case 1 
neutron spectra with $\lambda_{s}$ = 300 and $\delta$ = 0.20. Panel (b) is based on Case 2 
neutron spectra with $\lambda_{s}\to\infty$ and $\delta$ = 0.20.}
\end{figure}

\begin{figure}
   \centerline{\includegraphics[width=\textwidth]{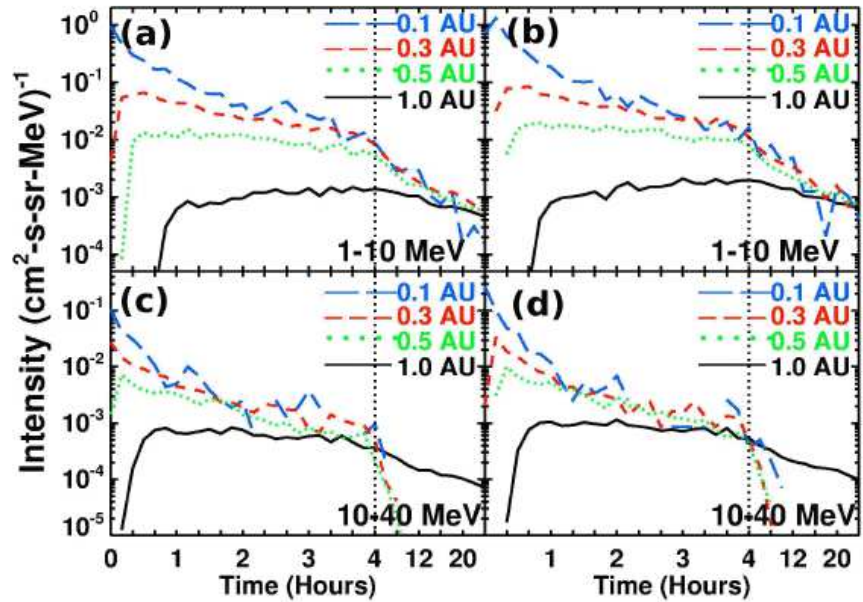}}
\caption{Proton intensities are shown for spacecraft at 0.1, 0.3, 0.5 and 1 AU  
following a central-meridian $\gamma$-ray flare with 10$^{32}$ accelerated protons 
$>$30 MeV. Panels (a) and (c) are based on Case 1 neutron spectra while Panel (b) 
and (d) are based on Case 2 neutron spectra. To the left of the black dotted line, 
the intensities are binned in 10-minute increments, and to the right of the dotted 
line the intensities are binned in 2-hour increments. The virtual spacecraft in this 
case are in the magnetically-well-connected sector spanning 45$^\circ$, so the 
black solid lines in this plot are the same as the cyan triangle lines in Figure 1.  
The mean statistical error on the 10-minute points increases closer to the Sun, and 
is $\sim$10\% on the 0.5 AU line, $\sim$20\% on the 0.3 AU line and $\sim$40\% on the 
0.1 AU line.} 
\end{figure}

As expected, these simulation results show that neutron-decay protons should be 
more easily detectable at heliocentric distances less than 1 AU.  The simulated 
spacecraft in Figure 4 are all in the magnetically-connected sector (45$^\circ$ 
clockwise from radial) but at different distances from the 
Sun, so the 1 AU results in the top portion of Figure 4 are the same as the 
results for that sector in Figure 1.  Within 0.5 AU heliocentric distance, 
neutron-decay protons peak within the first hour, and remain well above nominal 
background levels for a full day.  The peak fluxes at 0.1 AU, 0.3 AU and 0.5 AU 
are respectively $\sim$1000, $\sim$100 and $\sim$10 times higher than at 1 AU.   
The {\it Solar Orbiter} and {\it Solar Probe Plus} missions, which will approach to within 
62 and 9.5 solar radii of the Sun (respectively), will carry instrumentation that 
should detect neutron-decay protons and electrons in a number of events. 

Close to the Sun, a shock from the associated CME can overtake 
neutron-decay protons with energies of a few MeV or less if their radial transport 
is stalled by pitch-angle scattering (see Figure 2).  Thus neutron-decay protons 
may provide seed particles for acceleration by CME-driven shocks, as suggested by 
\inlinecite{fel10}.

\section{Conclusions}

The current solar cycle should present an excellent opportunity to observe neu\-tron-decay 
protons with multiple spacecraft over different heliographic longitudes and distances 
from the Sun.  Our simulation suggests that these neutron-decay 
protons are very rapidly distributed over more than 180$^\circ$ in longitude and should 
be detectable by current instrumentation at more than one longitude following large 
$\gamma$-ray events, provided the pre-event energetic particle background is sufficiently 
quiet. Spacecraft closer to the Sun such as {\it MESSENGER}, and eventually {\it Solar Probe Plus} and {\it Solar Orbiter}, will observe even higher intensities of neutron-decay protons.  While the decay protons and electrons with energies greater than a few MeV quickly escape into the heliosphere, lower-energy protons linger close to the Sun, where they are available to be accelerated by a CME shock into high-energy particles.

\begin{acks}
This work was supported by NASA under grants NNX08AI11G and NNX11AO75G, 
and subcontract SA2715-26309 from UC Berkeley under NASA 
contract \linebreak NAS5-03131. We appreciate the 
assistance of R. A. Leske and C. M. S. Cohen, and discussions with R. Murphy. 
We also thank the anonymous reviewer for a number of helpful comments and suggestions.  
\end{acks}

\end{article}

\end{document}